\documentclass[aps,pra,twocolumn,preprintnumbers]{revtex4}
\usepackage{graphicx}
\usepackage{amssymb}
\usepackage{amsmath}
\newcommand{\be}{\begin{equation}}
\newcommand{\ee}{\end{equation}}
\newcommand{\bea}{\begin{eqnarray}}
\newcommand{\eea}{\end{eqnarray}}
\newcommand{\Eq}[1]{Eq.\,(\ref{#1})}
\newcommand{\Fig}[1]{Fig.\,\ref{#1}}
\newcommand{\Sec}[1]{Sec.\,\ref{#1}}

\newcommand{\EPeval}{e}

\newcommand{\EPrevec}{\mathbf{e}^{>}}
\newcommand{\EPlevec}{\mathbf{e}^{<}}

\newcommand{\sigmaS}{\ensuremath{\sigma_{\rm S}}}
\newcommand{\sigmaeps}{\ensuremath{\sigma_\varepsilon}}

\newcommand{\bx}{\mathbf{x}}

\begin{document}
\title{No exceptional precision of exceptional point sensors}
\author{W. Langbein}
\affiliation{School of Physics and Astronomy, Cardiff University, Cardiff CF24 3AA,
United Kingdom}
\begin{abstract}
Recently, sensors with resonances at exceptional points (EPs) have been suggested to have a vastly improved sensitivity due to the extraordinary scaling of the complex frequency splitting of the $n$ initially degenerate modes with the $n$-th root of the perturbation. We show here that the resulting quantum-limited signal to noise at EPs is proportional to the perturbation, and comparable to other sensors, thus providing the same precision. The complex frequency splitting close to EPs is therefore not suited to estimate the precision of EP sensors. The underlying reason of this counter-intuitive result is that the mode fields, described by the eigenvectors, are equal for all modes at the EP, and are strongly changing with the perturbation.
\end{abstract}
%
%
\date{\today}
\maketitle
\section{Introduction}
In recent literature, following a first proposal \cite{WiersigPRL14}, optical sensors operating at so-called exceptional points (EP) have been discussed theoretically \cite{ZhangSR15,WiersigPRA16,RenOL17} and also recently experimentally \cite{ChenN17,HodaeiN17}.
The main attraction of EP sensors is that the splitting of the complex eigenfrequencies of the $n$ initially degenerate modes is scaling with the $n$-th root of the perturbation strength $\varepsilon$, which is extraordinary compared to the linear scaling with $\varepsilon$ for other sensors. As a result, the rate of change of the complex frequency splitting with $\varepsilon$ can be arbitrarily large for small perturbations, and is proportional to $1/\sqrt{\varepsilon}$ in the most studied case with $n=2$. 
The resulting enhanced complex frequency splitting for a given perturbation was claimed be exploitable for ultrasensitive mass sensing \cite{WiersigPRL14}, and to pave the way for sensors with unprecedented sensitivity \cite{ChenN17}. In this respect, it is important to note that the term ``sensitivity'' is an ambiguous quantity, as it can either refer to a transduction coefficient of the sensor from the quantity to be measured to some intermediate output quantity  (such as the frequency splitting in the present case), or to the smallest measurable change of the input quantity given by the noise of the sensor output. The latter is the precision of the measurement and is well defined. Surprisingly, the precision was not evaluated for EP sensors in all these works. 

Now, considering that the actual physical mechanism of the perturbation for the optical sensors considered in \cite{WiersigPRL14,ZhangSR15,WiersigPRA16,RenOL17,ChenN17,HodaeiN17} is the field change created by the interaction of the mode field with the polarisability of the perturbation, one would expect \cite{DoostPRA14} that in the limit of small perturbations, the resulting measurable effect, the change of the field, is linear with the perturbation strength $\varepsilon$, which is proportional to the change of the permittivity by the perturbation. 
It is therefore worth while to investigate the significance of the complex frequency splitting in terms of the measurable quantities which have a defined noise limit. While we might think that the measurable quantities would be proportional to the frequency splitting, based on our experience with well separated resonances having a linewidth smaller than the splitting, the situation is less clear for overlapping resonances with a linewidth larger than the splitting, which is the case for the EP sensors.         

The paper is organized as follows. In \Sec{sec:Ham} we introduce the mathematical description of the sensor in terms of its Hamiltonian and its eigenstates, and calculate its response in \Sec{sec:Sig}, giving general expressions in \Sec{sec:Sensig}. The signals of EP and DP sensors when perturbed from their degeneracy point are discussed in \Sec{sec:SensigEPDP}, and the precision at finite perturbation is shown in \Sec{sec:SensPrec}. In the appendix we analyse published data from an EP sensor to determine its precision.  

\section{Hamiltonian and Eigenstates}\label{sec:Ham}

To discuss these conflicting expectations for the scaling of the signal field with the perturbation, let us use the simple case of two degenerate states at the EP as discussed in the literature \cite{HeissJPA12, WiersigPRL14, ChenN17}. The corresponding 2x2 Hamiltonian can be written as 
\be H_0 = 
\begin{pmatrix}
	E_0 & A_0 \\
	0 & E_0 \\ 
\end{pmatrix}
\label{eqn:H0} \ee
with the degenerate state frequencies $E_0$ and the off-diagonal coupling frequency $A_0$. For $A_0\neq0$, this matrix has only one eigenvalue $E_0$ and right eigenvector $(1,0)^\top$, i.e. its two eigenvalues are degenerate with equal eigenvectors. For $A_0=0$ instead, a case known \cite{WiersigPRL14} as ``diabolic'' point (DP), the two degenerate eigenvalues have orthogonal right eigenvectors $(1,0)^\top$ and $(0,1)^\top$. This observation gives us already a clue of the difference of the response to a perturbation between EP and DP -- the degeneracy of the eigenvectors at the EP is lifted together with the degeneracy of the eigenvalues, leading to a quadratic behaviour of the measurable signal in the energy splitting, thus recovering the linearity with perturbation strength. But lets see this via a mathematical derivation, where we introduce the perturbed Hamiltonian as
\be H = H_0 + \varepsilon H_1, \quad \mbox{with} \quad H_1= 
\begin{pmatrix}
	 0 & A_1 \\
	B_1 & 0 \\ 
\end{pmatrix}
\label{eqn:H1} \ee
with the perturbation strength $\varepsilon>0$, and the perturbation matrix $H_1$ given by generally complex coupling frequencies $A_1\neq0$ and $B_1\neq0$. For simplicity, following \cite{WiersigPRA16}, we have set the diagonal elements of $H_1$ to zero as they do not provide the square-root frequency splitting at the EP. The resulting eigenvalues are 
\be \EPeval_{\pm} = E_0 \pm \sqrt{\varepsilon A_0B_1 + \varepsilon^2A_1B_1} = E_0 \pm \Delta\,,
\label{eqn:eval} \ee
having the left ($\EPlevec_\pm$) and right ($\EPrevec_\pm$) eigenvectors 
\be 
\EPlevec_\pm= \frac{1}{\sqrt{2}}
\left(\frac{1}{v} , \pm 1\right) \quad 
\EPrevec_\pm= \frac{1}{\sqrt{2}}
\begin{pmatrix}
	v \\
	\pm 1 \\ 
\end{pmatrix}
\label{eqn:evec}\ee
where $v=\Delta/(\varepsilon B_1)$. They are normalized according to $\EPlevec_\pm\EPrevec_\pm=1$ suited for modal decomposition. We see that for an EP, the eigenvectors depend on $\varepsilon$, becoming parallel for vanishing $\varepsilon$. 

\section{Hamiltonian dynamics and sensing signals}\label{sec:Sig}

Measuring the system involves exciting the system, and detecting its response. Changes in the response with the perturbation are then used to determine the perturbation. In \cite{WiersigPRL14, ChenN17} the basis of the Hamiltonian are the clockwise and counterclockwise propagating optical modes of a microtoroid, which can be separately excited and detected by evanescent coupling to a single mode fibre in the two propagation directions in the fibre. These modes correspond to the first and second element of the eigenvectors, respectively, in the above formulation.

Detecting the field, for example using heterodyne detection \cite{DemtroderBook98}, the noise in the detection is quantum-noise limited to a given field uncertainty, which is due to a Heisenberg uncertainty relation between the real and imaginary part of the field in the rotating wave picture. This limit is called the standard quantum limit and is discussed for example in \cite{BraginskyBook92}. It is consistent with the shot-noise-limit for intensity measurements. In the experiments reported in \cite{ChenN17,HodaeiN17}, a tuneable single frequency laser was used as excitation source and the transmitted and reflected power was measured using photodetectors. This measurement is quantum limited by the photon shot noise, which is equivalent to the standard quantum limit for the measurement of the optical electric field.  
 
Therefore, we can evaluate the quantum limited precision using the change of the detected field due to the perturbation. The absolute sensitivity will depend on the excitation field amplitude and calculating it is not required for the comparison of the EP and DP sensors. We will work in the rotating wave picture and omit the complex conjugate part required to describe real fields, for brevity.

\subsection{Sensor signal}\label{sec:Sensig}

The Hamiltonian dynamics of the field $\mathbf{S}(t)$ of the system with an excitation $\bx(t)$ is given by
\be
\partial_t\mathbf{S}(t)= iH\mathbf{S}(t)+\bx(t)\,.
\label{eqn:SdefG}\ee
which for an initialy unexcited system yields
\be
\mathbf{S}(t)= \int_{-\infty}^t \exp\left(iH(t-t')\right) \bx(t') dt'\,.
\label{eqn:Sdyn}\ee
Using the modal decomposion of $H$, this simplifies to
\be
\mathbf{S}(t)=\sum_\pm \EPrevec_\pm\EPlevec_\pm \int_{-\infty}^t \exp\left(ie_\pm (t-t')\right) \bx(t') dt'\,,
\label{eqn:SdefD}\ee
which for $\bx(t)=\delta(t)\bx_0$ yields 
\be
\mathbf{S}(t)=\theta(t)\hat{S}\bx_0=\theta(t)\sum_\pm \EPrevec_\pm\exp\left(ie_\pm t\right)\EPlevec_\pm \bx_0\,,
\label{eqn:Sdef}\ee
with the Heaviside function $\theta(t)$. Using the expressions for the eigenvalues in \Eq{eqn:eval} and the eigenvectors in \Eq{eqn:evec}, we find
\be \hat{S}=\exp\left(iE_0t\right)
\begin{pmatrix}
	\cos(\Delta t) & i\sin(\Delta t)v \\
	i\sin(\Delta t)/v & \cos(\Delta t) \\
\end{pmatrix}\,.
\label{eqn:Shat}\ee

To evaluate the frequency domain response, we Fourier-transform the time-domain response \Eq{eqn:Shat} into the angular frequency domain using 
$\tilde{S}(\omega)=\int S(t)\exp(-i\omega t)dt$, and find  
\be \tilde{\hat{S}}=\frac{1}{2}\begin{pmatrix}
	p_+ + p_- & (p_+ - p_-)v \\
	(p_+ - p_-)/v & p_+ + p_-\\
\end{pmatrix}\,,
\label{eqn:Shatf} \ee
with $p_\pm=i/(\Omega\pm\Delta)$, using $\Omega=E_0-\omega$. 

\subsection{Sensor precision at an EP or a DP}\label{sec:SensigEPDP}

At an EP, subtracting the response for $\varepsilon=0$ yields  
\be
\hat{S}_\varepsilon^{\rm E}=\exp\left(iE_0t\right)
\begin{pmatrix}
	\cos(\Delta t)-1 & i(v\sin(\Delta t)-A_0t) \\
	i\sin(\Delta t)/v & \cos(\Delta t)-1 \\
\end{pmatrix}\,.
\label{eqn:EPSs}\ee
Now developing in orders of $\alpha=\sqrt{\varepsilon B_1/A_0}$, 
\bea \Delta= \alpha A_0 + {\cal O}(\alpha^3) \quad , \quad  \frac{1}{v}=\alpha+{\cal O}(\alpha^3)\nonumber\,,\quad \mbox{and} \\
v\sin(\Delta t)=A_0t+\left(\frac{A_0A_1}{B_1}t-\frac{A_0^3}{6}t^3\right)\alpha^2+{\cal O}(\alpha^4)\,, \nonumber
\label{eqn:EPdev} \eea
we find using $\tau=A_0 t$
\be
\hat{S}_\varepsilon^{\rm E}=\alpha^2\exp\left(iE_0t\right)
\begin{pmatrix}
	-\tau^2/2 & i\left(\frac{A_1}{B_1}\tau-\frac{\tau^3}{6}\right) \\
	i\tau & -\tau^2/2 \\
\end{pmatrix}
 + {\cal O}(\alpha^4)\,, 
\label{eqn:EPSdev}\ee
which in lowest order in $\alpha$ is
\be
\hat{S}_\varepsilon^{\rm E}\approx\varepsilon t \exp\left(iE_0t\right)
\begin{pmatrix}
	-\tau B_1/2 &  iA_1-iB_1\tau^2/6\\
	iB_1 & -\tau B_1 /2 \\
\end{pmatrix}\,,
\label{eqn:EPSdevlo}\ee
a signal proportional to the perturbation $\varepsilon$, as expected from the initial physical argument. 

The frequency domain response corresponding to \Eq{eqn:EPSs} is given by
\be \tilde{\hat{S}}_\varepsilon^{\rm E}=\frac{1}{2}\begin{pmatrix}
	p_+ + p_- -2p_0 & (p_+ - p_-)v - 2iA_0p_0^2 \\
	(p_+ - p_-)/v & p_+ + p_- -2p_0 \\
\end{pmatrix}\,,
\label{eqn:EPSfs} \ee
with $p_0=i/\Omega$. Developing in $\alpha$ yields in lowest order
\be  \tilde{\hat{S}}_\varepsilon^{\rm E} \approx 
\frac{\varepsilon}{\Omega^2}\begin{pmatrix}
	A_0 B_1/\Omega & -i(A_1+A_0^2B_1/\Omega^2) \\
	-iB_1  & A_0 B_1/\Omega \\
\end{pmatrix}\,,
\label{eqn:EPSfsdev} \ee
again scaling proportional to the perturbation $\varepsilon$.  

We thus note that for all excitation - detection cases and small perturbations, the measurable signal field from the EP sensor is proportional to the perturbation, despite the complex frequency difference scaling with the square root of the perturbation.

Now, for comparison, lets consider the situation for a DP sensor, given by $A_0=0$. The eigenvectors, given by \Eq{eqn:evec}, are now independent of the perturbation strength $\varepsilon$, and subtracting the signal for $\varepsilon=0$ we have
\be
\hat{S}_\varepsilon^{\rm D}=\exp\left(iE_0t\right)
\begin{pmatrix}
	\cos(\Delta t)-1 & i\sin(\Delta t)v \\
	i\sin(\Delta t)/v & \cos(\Delta t)-1 \\
\end{pmatrix}\,,
\label{eqn:DPSs}\ee
and developing in $\Delta=\varepsilon\sqrt{A_1B_1}$ yields in lowest order
\be
\hat{S}_\varepsilon^{\rm D}\approx\exp\left(iE_0t\right)
\begin{pmatrix}
	-\Delta^2 t^2/2 & i\Delta t v \\
	i\Delta t/v & -\Delta^2 t^2/2 \\
\end{pmatrix}\,,
\label{eqn:DPSsdev}\ee
which written in $\varepsilon$ becomes
\be
\hat{S}_\varepsilon^{\rm D}\approx\varepsilon t\exp\left(iE_0t\right)
\begin{pmatrix}
	-\varepsilon A_1B_1 t/2 & i A_1 \\
	i B_1 & -\varepsilon A_1B_1 t/2 \\
\end{pmatrix}\,,
\label{eqn:DPSsdeve}\ee
again proportional to $\varepsilon$ in lowest order. Notably, the off-diagonal elements linear in $t$ are equal to the EP result, while the diagonal elements are in lowest order proportional to $\varepsilon^2$, different from the EP result \Eq{eqn:EPSdevlo}. In spectral domain, we find    
\be  \tilde{\hat{S}}^{\rm D}_\varepsilon=\frac{1}{2}\begin{pmatrix}
	p_+ + p_--2p_0 & (p_+ - p_-)v \\
	(p_+ - p_-)/v & p_+ + p_- -2p_0 \\
\end{pmatrix}
\label{eqn:DPSfs} \ee

and developing in $\Delta$ yields in lowest order
\bea  \tilde{\hat{S}}_\varepsilon^{\rm D}&\approx&\frac{\Delta}{\Omega^2}\begin{pmatrix}
	i\Delta/\Omega & -iv \\
	-i/v & i\Delta/\Omega \\
\end{pmatrix}\\\nonumber
&=&\frac{\varepsilon}{\Omega^2}\begin{pmatrix}
	i\varepsilon A_1B_1/\Omega & -iA_1 \\
	-iB_1 & i\varepsilon A_1B_1/\Omega \\
\end{pmatrix}
\label{eqn:DPSfsdev} \eea
Also here, the off-diagonal elements are equal to each other, and identical to the EP result apart from the $\Omega^{-4}$ term in \Eq{eqn:EPSfsdev}. 
\begin{figure}
	\includegraphics*[scale=0.60]{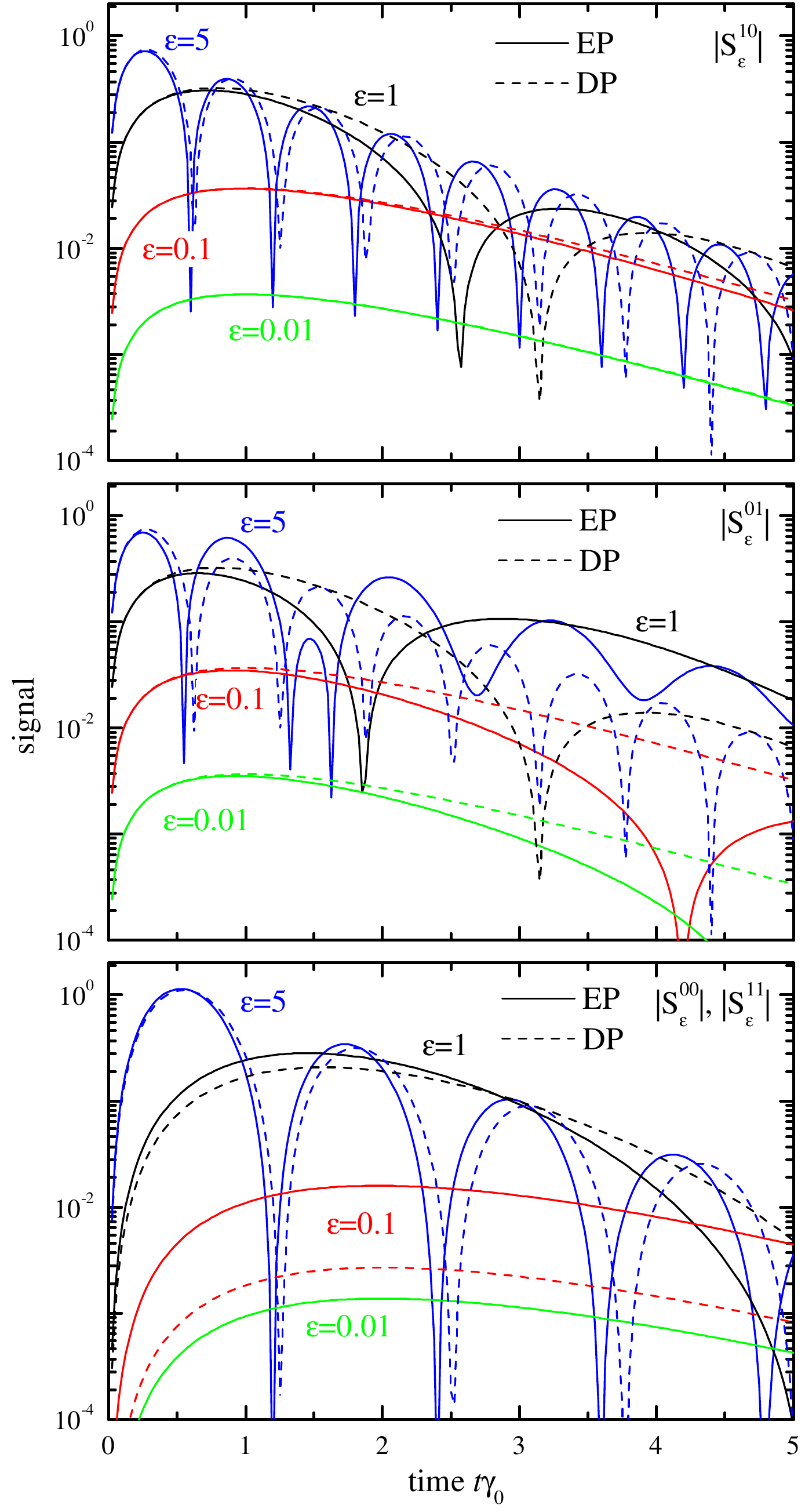}
	\caption{Time-domain sensing signal amplitudes for sensors at an EP (solid lines, $A_0=\gamma_0/2$, \Eq{eqn:EPSs}) or DP (dashed lines, $A_0=0$, \Eq{eqn:DPSs}) for different perturbation strengths $\varepsilon$ as labeled and color-coded, using $A_1=B_1=\gamma_0$. Top: $|S_\varepsilon^{10}|$,  Middle: $|S_\varepsilon^{01}|$; Bottom: $|S_\varepsilon^{00}|$ and $|S_\varepsilon^{11}|$.}
	\label{fig:Time}
\end{figure}

\begin{figure}
	\includegraphics*[scale=0.60]{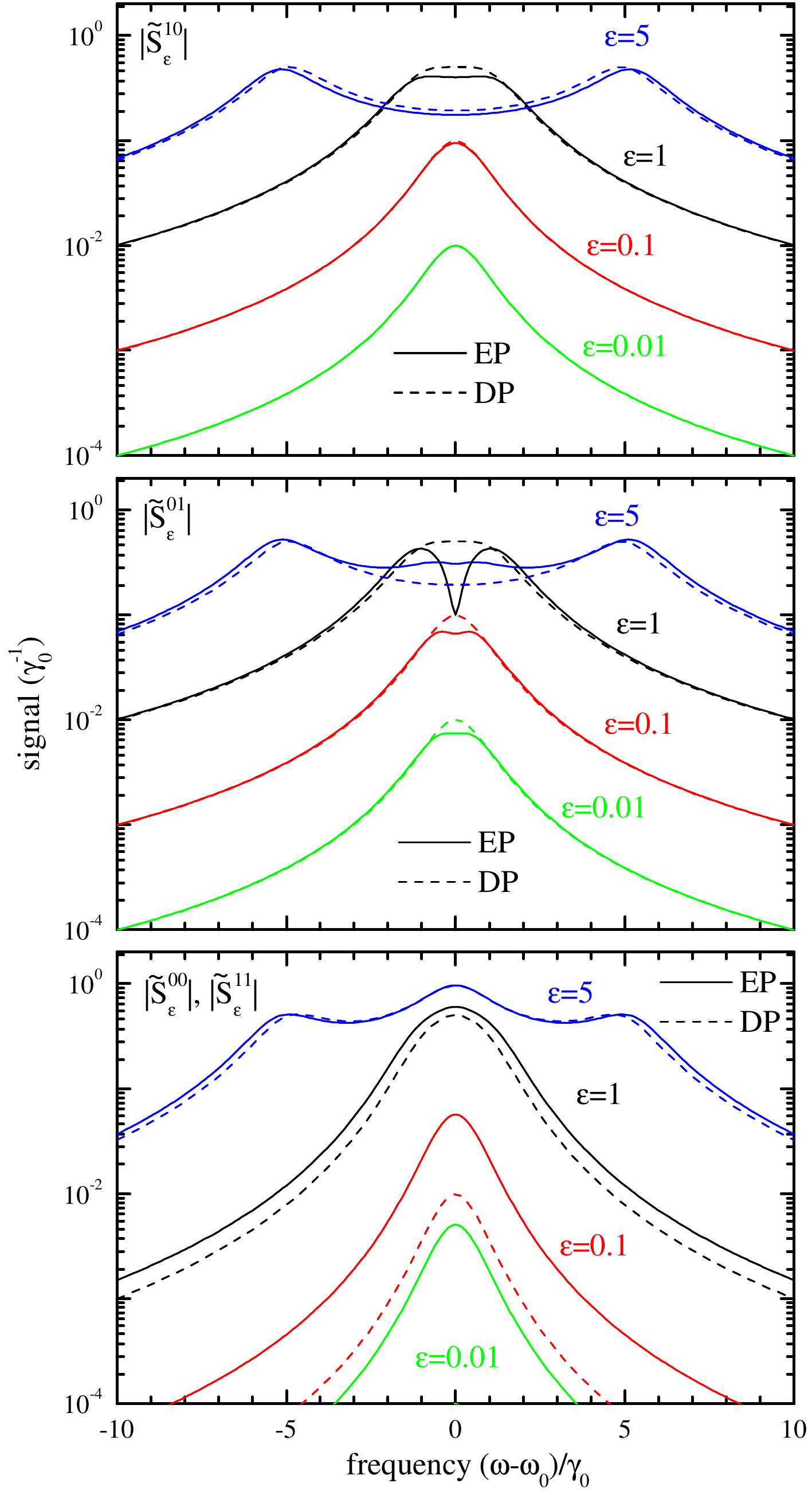}
	\caption{As \Fig{fig:Time}, but in frequency-domain, using \Eq{eqn:EPSfs} and \Eq{eqn:DPSfs}.}
	\label{fig:Freq}
\end{figure}

Note that the response of the system to an arbitrary excitation pulse in time domain is given by the convolution of the response for delta-excitation, the Green's function, given by \Eq{eqn:Sdef}, with the excitation pulse. Furthermore, in spectral domain, the response is given by the product of the spectral response function, \Eq{eqn:EPSfs}, and the excitation spectrum. The peak of the spectral response function therefore gives the highest achievable spectral response for any excitation spectrum.

For illustration, we show the time-domain signals in \Fig{fig:Time}, for sensors at an EP, given by \Eq{eqn:EPSs}, or a DP, given by \Eq{eqn:DPSs}, using perturbation strengths $\varepsilon=5$, 1, 0.1, and 0.01. The corresponding frequency domain signals are given in \Fig{fig:Freq}, according to \Eq{eqn:EPSfs} for an EP and \Eq{eqn:DPSfs} for a DP sensor.

We split the complex frequency $E_0=\omega_0+i\gamma_0$ into the real frequency $\omega_0$ and the damping $\gamma_0$, take $A_0=\gamma_0/2$, which is the maximum possible due to the physical constrain of a dissipative system \cite{WiersigPRA16}, and use $A_1=B_1=\gamma_0$, which provides a lossless scattering between the two modes. 

The off-diagonal signal $|S_\varepsilon^{10}|$ (row 1, column 0 of $S_\varepsilon$) shown in the top panels, is equal for EP and DP apart from the slightly larger frequency splitting $\Delta$ for the EP (see \Eq{eqn:eval}), which is directly observable in \Fig{fig:Freq}, and results in a faster temporal beating period seen in \Fig{fig:Time}.

The off-diagonal signal $|S_\varepsilon^{01}|$ shown in the middle panels is the same as $|S_\varepsilon^{10}|$ for the DP but is different for the EP, where it is not background free. It shows an additional contribution around zero detuning in frequency domain, which interferes destructively with the main signal, leading to a strong suppression for $\varepsilon=1$ at zero detuning.

The diagonal signals $|S_\varepsilon^{00}|$, $|S_\varepsilon^{11}|$ are equal and shown in the bottom panels, and are not background free for both EP and DP. They are created only by the changing temporal dynamics of the signal, providing an initial quadratic rise. However, since $\Delta$ is scaling for $\varepsilon \ll 1$ with $\sqrt{\varepsilon}$ for the EP, and with $\varepsilon$ for the DP, the resulting signal is scaling for $\varepsilon \ll 1$ as $\varepsilon$ for the EP, and as $\varepsilon^2$ for the DP. 

\begin{figure}
	\includegraphics*[scale=0.60]{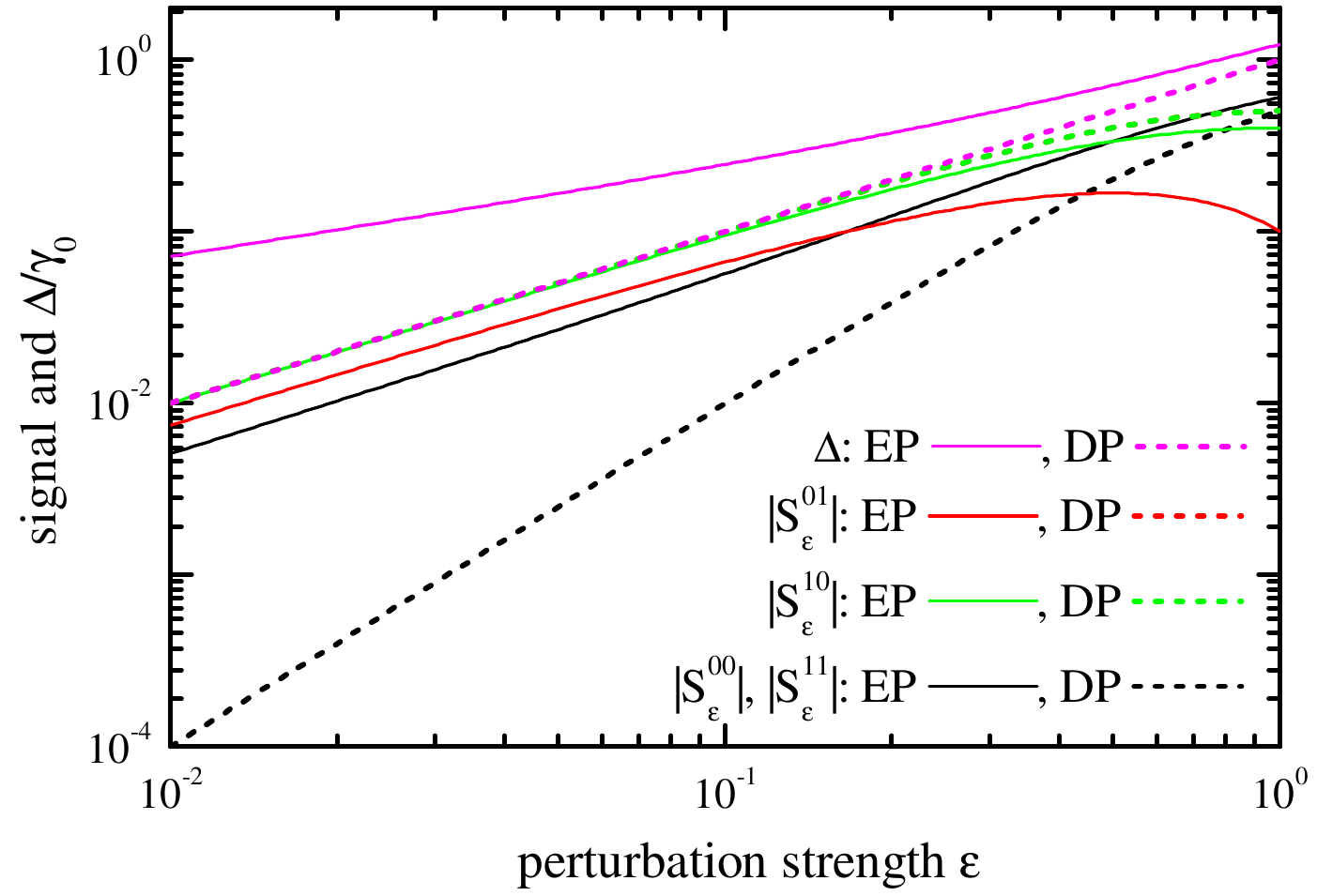}
	\caption{Frequency-domain sensing signal amplitudes and complex frequency splitting $\Delta$ for sensors at an EP (solid lines) or DP (dashed lines) at resonance ($\omega=\omega_0$), as function of the perturbation strength $\varepsilon$. Other parameters as in \Fig{fig:Time}.}
	\label{fig:Comp}
\end{figure}

We note that for the off-diagonal signals the time-domain peak is scaling as $\gamma_0^{-1}$, and thus with the quality factor of the modes given by $\omega_0/(2\gamma_0)$. The frequency domain scaling acquires an extra factor $\gamma_0^{-1}$ provided by the changing time duration, for all signals. The EP off-diagonal signal $S_\varepsilon^{{\rm E},10}$ has an additional component scaling as $\gamma_0^{-3}$ in time domain, which is interfering with the main component giving rise to additional features. For the diagonal signals the time-domain peak is scaling as $\gamma_0^{-2}$. Therefore, for all signal components, smaller damping results in higher signals. In this respect, we note that in the graphs presented we used the same loss rate $\gamma_0$ for both EP and DP sensor. However, due to the physical constrain \cite{WiersigPRA16}
$2\gamma_0\ge|A_0|$ for a passive EP sensor, the loss rate is typically higher for an EP sensor than for a DP sensor, which is also observed in simulations \cite{WiersigPRA16} and experiment \cite{ChenN17}. Introducing optical gain into the system to reduce the loss can be done in EP and DP systems, and can enhance the signal of both. However, one has to keep in mind that gain processes introduce additional noise in the signal, and will not be further discussed here. 

The dependence of the complex frequency splitting and the frequency domain signals on the perturbation is shown in \Fig{fig:Comp} for both EP and DP sensors, at resonance. We find that all signals are scaling for small perturbations linear with the perturbation, except $|S_\varepsilon^{00}|$ and $|S_\varepsilon^{11}|$ for the DP sensor which scales quadratically. The highest signal is provided by $|S_\varepsilon^{10}|$ of the DP sensor, nearly matched by $|S_\varepsilon^{10}|$ of the EP sensor. Importantly, while for the DP sensor the frequency splitting $\Delta$ is scaling proportional to the signal, for the EP sensor this splitting is scaling differently than the signal, being proportional to $\sqrt{\varepsilon}$ for small $\varepsilon$. This clearly shows that for EP sensors the complex frequency splitting is not suited to estimate the precision. 

This finding is consistent with an analysis of the sensing precision in the data provided in \cite{ChenN17} (see Appendix \ref{sec:chendata}), which shows that the data taken for the EP have higher noise in measuring the perturbation compared to the data taken for the DP.  

\subsection{Sensor precision at finite perturbations}\label{sec:SensPrec}

To determine the sensor precision at finite perturbation, we analyze here the change of the signal due to a change of $\varepsilon$ at finite values of $\varepsilon$, i.e. detuned from the EP or DP. This change is given by the responsivity of the sensor, defined as the derivative of $\hat{S}$ (see \Eq{eqn:Shat}) with respect to $\varepsilon$, 
\be \label{eqn:TimeDif} \hat{D}_\varepsilon=\frac{d\hat{S}}{d\varepsilon}=\exp\left(iE_0t\right)\begin{pmatrix}
	d^{00} & d^{10}\\
	d^{01} & d^{11}\\
\end{pmatrix} \ee  
where 
\bea
d^{00}&=& d^{11}=-B_1 t \left(\frac{A_0}{2}+\varepsilon A_1\right)\frac{\sin(\Delta t)}{\Delta}\,,\\
d^{01}&=&i\left(\frac{A_0}{2\varepsilon}+A_1\right)\left(t\cos(\Delta t)+\frac{\sin(\Delta t)}{\Delta}\right)\nonumber\\&&-\frac{i \Delta \sin(\Delta t)}{\varepsilon^2 B_1}\,,\\
d^{10}&=&i\left(\frac{A_0}{2}+\varepsilon A_1\right)\left(t\cos(\Delta t)-\frac{\sin(\Delta t)}{\Delta}\right)\frac{\varepsilon B_1^2}{\Delta^2}\nonumber\\&&+i B_1\frac{\sin(\Delta t)}{\Delta}\,.
\eea

\begin{figure}
	\includegraphics*[scale=0.60]{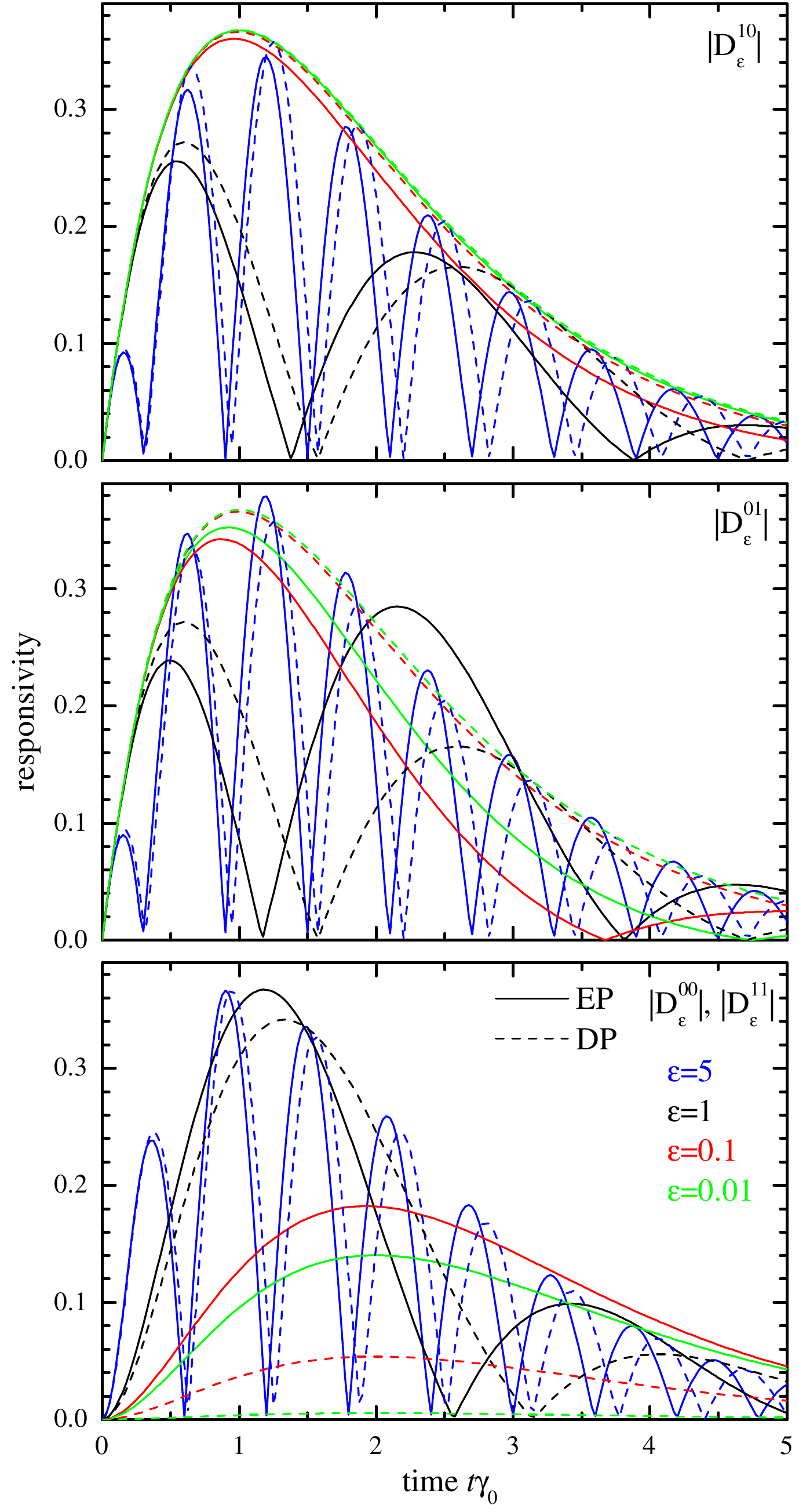}
	\caption{Responsivities according to \Eq{eqn:TimeDif} at finite $\varepsilon$ for sensors at an EP (solid lines, $A_0=\gamma_0/2$) or DP (dashed lines, $A_0=0$) for different perturbation strengths $\varepsilon$ as labelled and color-coded, using $A_1=B_1=\gamma_0$. Top: $|D_\varepsilon^{10}|$,  Middle: $|D_\varepsilon^{01}|$; Bottom: $|D_\varepsilon^{00}|$ and $|D_\varepsilon^{11}|$.}
	\label{fig:TimeDif}
\end{figure}

A quantum limited noise \sigmaS\ of the signal then results in a quantum limited precision of the sensor for changes in $\varepsilon$ given by
\be \label{eqn:DifPrec} \sigmaeps^{nm} = \sigmaS/|D_\varepsilon^{nm}|\,, \ee
with the indices $n,m \in \{0,1\}$ selecting the detected signal component. 
Examples of the responsitivities $|D_\varepsilon^{nm}|$ are given in \Fig{fig:TimeDif} using parameters as in \Fig{fig:Time}. We find that for $\varepsilon \ll 1$, the responsitivities have a stable amplitude versus $\varepsilon$, as expected for a signal scaling linear with $\varepsilon$. An exception are the diagonal components for the DP, which scale quadratic with $\varepsilon$, thus showing a responsivity proportional to $\varepsilon$ in this regime. For larger $\varepsilon$, oscillations versus time are present, reflecting the significant frequency splitting. Notably the maximum responsitivities are similar to ones in the $\varepsilon \ll 1$ regime.

For detection in frequency domain, we determine equivalently the derivative of $\tilde{\hat{S}}$ (see \Eq{eqn:Shatf}) versus $\varepsilon$, 
\be \label{eqn:FreqDif} \tilde{\hat{D}}_\varepsilon=\frac{d\tilde{\hat{S}}}{d\varepsilon}=\frac{1}{2}\begin{pmatrix}
	\tilde{d}^{00} & \tilde{d}^{10} \\
	\tilde{d}^{01} & \tilde{d}^{11}\\
\end{pmatrix} \ee  
where 
\bea
\tilde{d}^{00}&=& \tilde{d}^{11}=i\frac{B_1}{\Delta}\left(\frac{A_0}{2}+\varepsilon A_1\right)\left(p_+^2-p_-^2\right)\,,\\
\tilde{d}^{01}&=&\left(\frac{A_0}{2\varepsilon}+A_1\right)\left(i\left(p_+^2+p_-^2\right)+\frac{p_+ - p_-}{\Delta}\right)\nonumber\\&&-\Delta\frac{p_+-p_-}{\varepsilon^2 B_1}\,,\\
\tilde{d}^{10}&=&\left(\frac{A_0}{2}+\varepsilon A_1\right)\left(i\left(p_+^2+p_-^2\right)-\frac{p_+ - p_-}{\Delta}\right)\frac{\varepsilon B_1^2}{\Delta^2}\nonumber\\&&+B_1\frac{p_+-p_-}{\Delta}\,.
\eea

\begin{figure}
	\includegraphics*[scale=0.60]{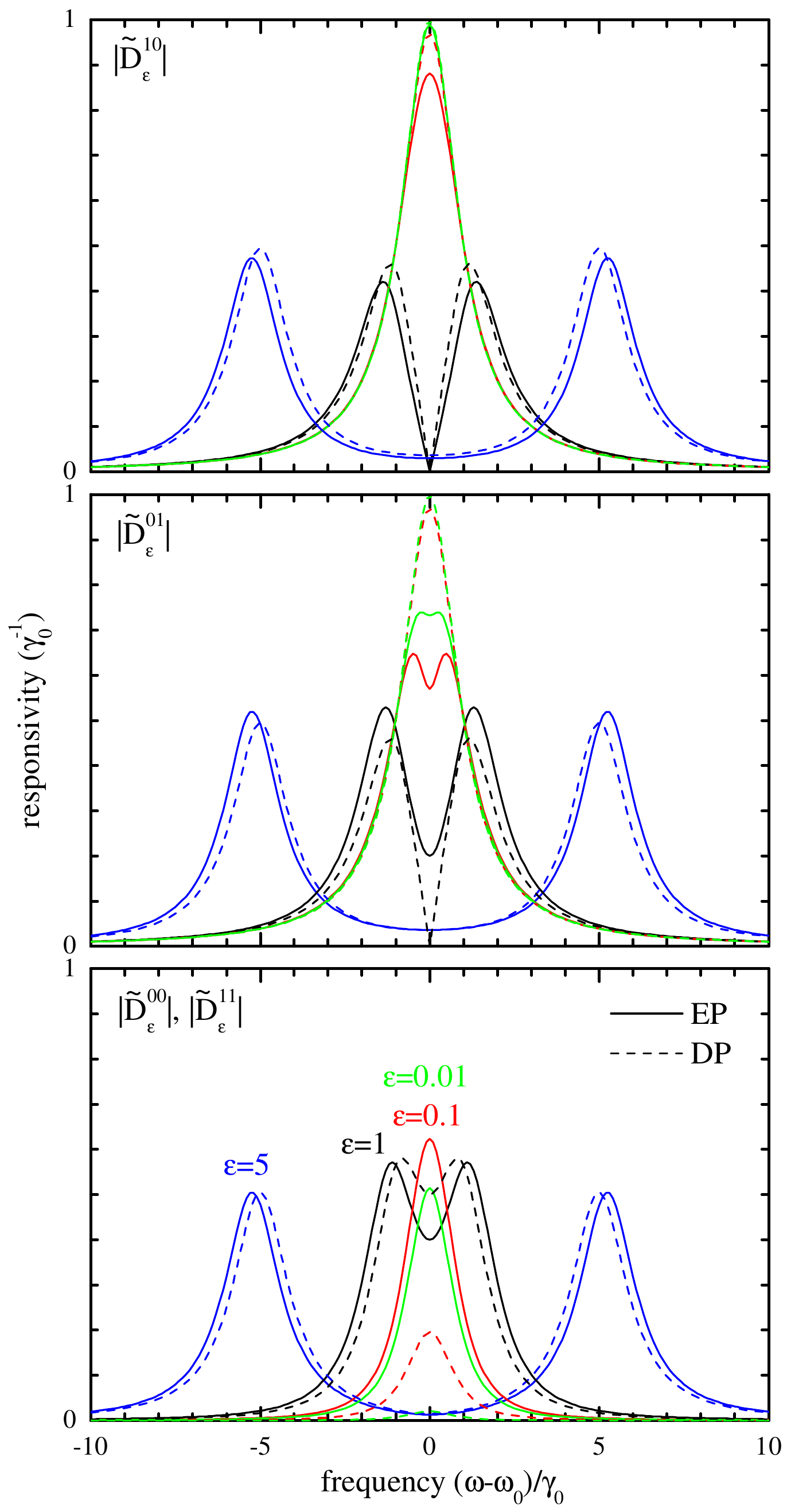}
	\caption{As \Fig{fig:TimeDif}, but in frequency domain, using \Eq{eqn:FreqDif} }
	\label{fig:FreqDif}
\end{figure}

The resulting frequency domain responsitivities $|\tilde{D}_\varepsilon^{nm}|$ for the parameters as in \Fig{fig:Freq} are given in \Fig{fig:FreqDif}. As expected, the responsivity is maximum when exciting the sensor on resonance. It is noticeable that, at resonance, the off-diagonal responsivity is twice as large for $\varepsilon \ll 1$ than for $\varepsilon \gg 1$, corresponding to cases of overlapping or separated resonances, respectively.
The quantum limited precision is still given by \Eq{eqn:DifPrec}, using the quantum limited noise in frequency domain.
The highest precision (i.e. the smallest $\sigma_\varepsilon^{nm}$) in frequency domain is achieved detecting the off-diagonal of a DP sensor, close to degeneracy (i.e. at $\varepsilon \ll 1$). Increasing $\varepsilon$ , the highest precision reduces, by a factor of two for $\varepsilon \gg 1$, for probing one of the resonances, see for example $\varepsilon=5$ in \Fig{fig:FreqDif} at $e_\pm \approx \omega_0 \pm 5\gamma_0$.   

Note that we evaluate the absolute value of the responsivity, which takes into account both amplitude and phase changes of the signal. The latter are dominating the sensor response at resonance for $\varepsilon \gg 1$, reflecting changes in the resonance frequency. 

We emphasize that resonance frequencies cannot be measured directly, but are deduced by fitting models of the response to measured fields or intensities. For example, for pulsed excitation, the time-dependent signal can be measured, and fitted with \Eq{eqn:Shat}. Exciting instead with a field of given frequency, the amplitude and phase of the signal at that frequency can be measured and fitted with \Eq{eqn:Shatf}.

\section{Conclusions}

In conclusion, we have demonstrated that the frequency splitting of the complex eigenfrequencies is not a suited measure for determining the precision of an EP sensor. Rather, the resulting detected signal has to be compared with the quantum noise limit of the detection, and with any additional technical noise. Such an analysis has not been presented in previous literature proposing EP sensors. From the analysis presented in the present work, it emerges that EP sensors do not provide the exceptional precision suggested by the frequency splitting scaling with the $n$-th root of the perturbation, but rather are comparable to other sensors, providing a signal field proportional to the perturbation strength, in agreement with expectations from first-order perturbation theory. We also provide explicit expressions for the sensor responsivity in time and frequency domain, which might prove helpful for sensor design.

\acknowledgments This work was supported by the EPSRC Grant 
EP/M020479/1, and the S\^er Cymru National Research Network in Advanced Engineering and Materials. The author acknowledges discussions with E.A. Muljarov.     

\appendix

\section{Analysis of data presented in Chen et al., Nature 548, 192 (2017) }
\label{sec:chendata}

\begin{figure}
	\includegraphics[width=\columnwidth]{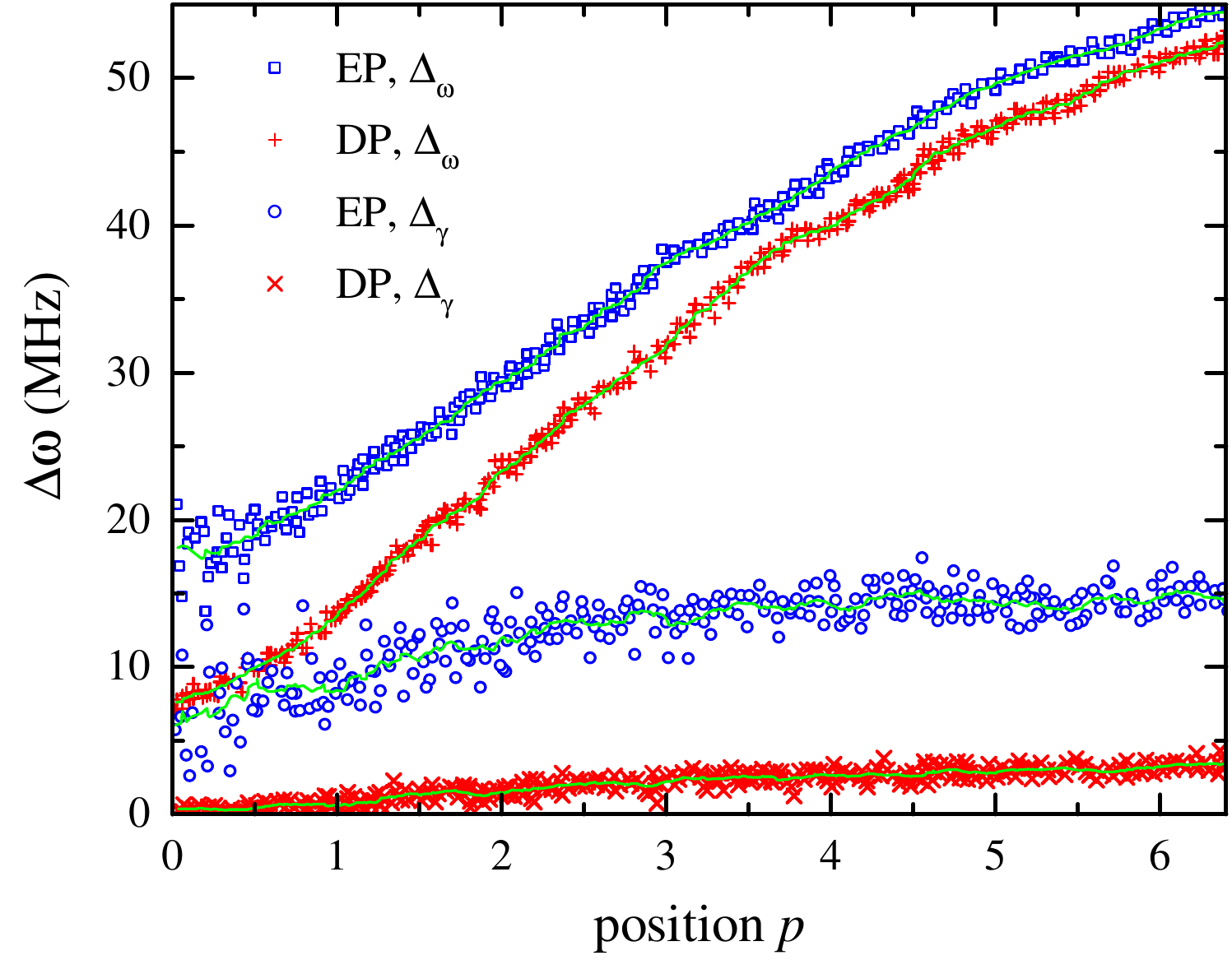}
	\caption{Measured frequency splitting $\Delta_\omega$ and linewidth difference $\Delta_\gamma$ for a sensor close to the EP or DP, as labelled, as function of the position of a target scatterer. Data digitized from Fig.2(a) in \cite{ChenN17}. The green lines are 10 point averages.}
	\label{fig:ChenFig2a}
\end{figure}

In the two experimental works on EP sensors \cite{ChenN17,HodaeiN17}, the precision of the investigated sensors was not reported. In order to investigate the precision, we therefore analyze the experimental data shown in \cite{ChenN17}, to extract the precision of the investigated sensor at the EP and DP point. The relevant data are shown in Fig.2(a) of \cite{ChenN17}, and a graph with the data digitized from this figure is shown in \Fig{fig:ChenFig2a}. The difference in linewidth $\Delta_\gamma$ and the difference in frequency $\Delta_\omega$ of the two modes are given as function of the position of a target scatter in arbitrary units. It is unclear in \cite{ChenN17} how this position relates to the distance of the fibre tip from the toroidal resonator used in the experiment. We will call this position $p$ in the following and treat it as unitless.   

\begin{figure}
	\includegraphics[width=\columnwidth]{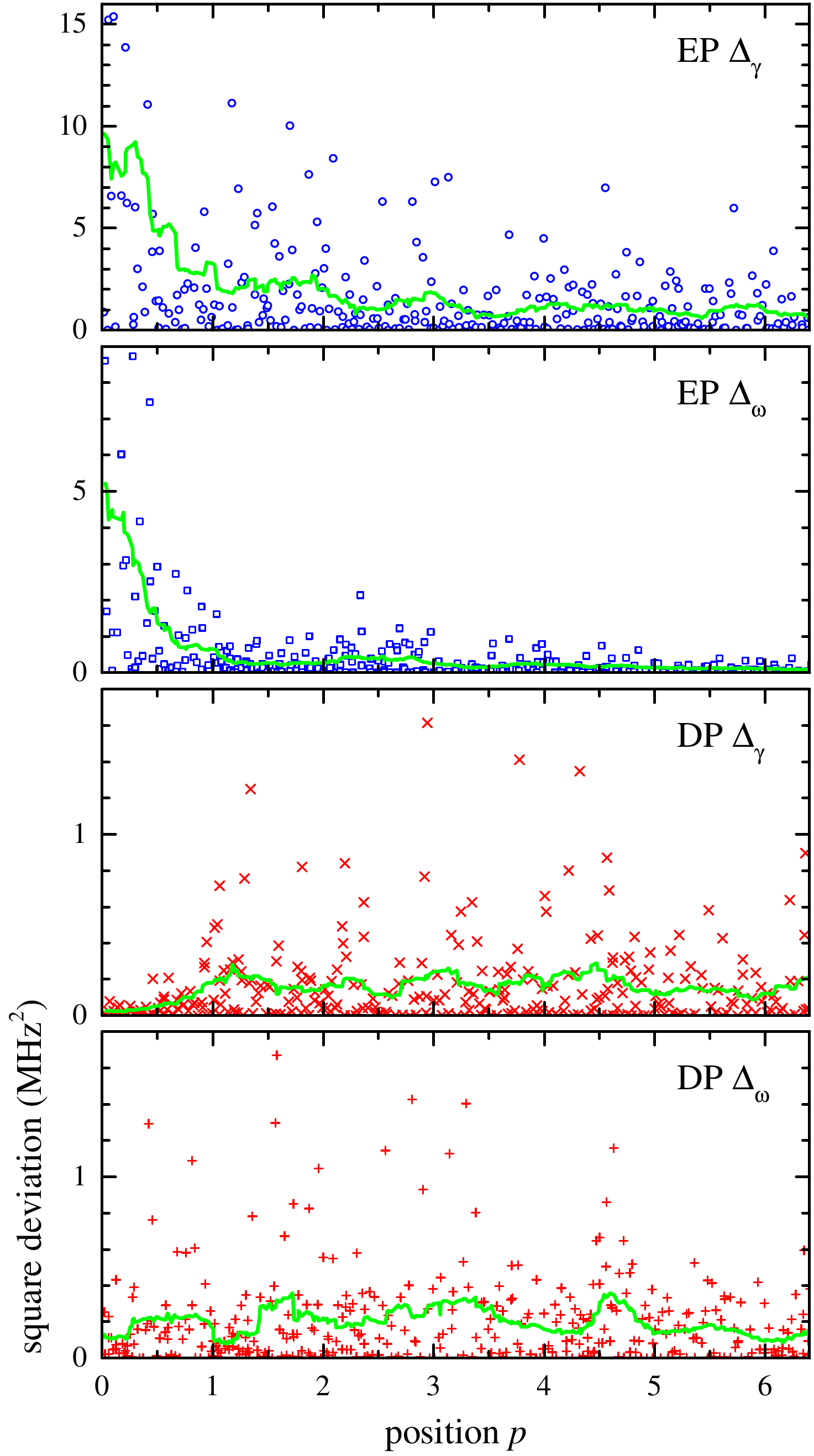}
	\caption{Square deviation between 10-point average and data for the measured frequency splitting $\Delta_\omega$ linewdith difference $\Delta_\gamma$ for a sensor close to the EP or DP, as labelled, as function of the position of a target scatterer in arbitrary units $p$. The green lines are 20-point averages.}
	\label{fig:SquareDev}
\end{figure}

In order to determine the precision of the sensing, we first have to determine the noise in the measurements, and then translate this to the corresponding noise in $p$. To determine the noise in the measured $\Delta_\omega$ and $\Delta_\gamma$, we calculate a 10-point adjacent average of the data points (see green lines in \Fig{fig:ChenFig2a}), and determine the square deviation of the data from this average, as shown in \Fig{fig:SquareDev}.

\begin{figure}
	\includegraphics[width=\columnwidth]{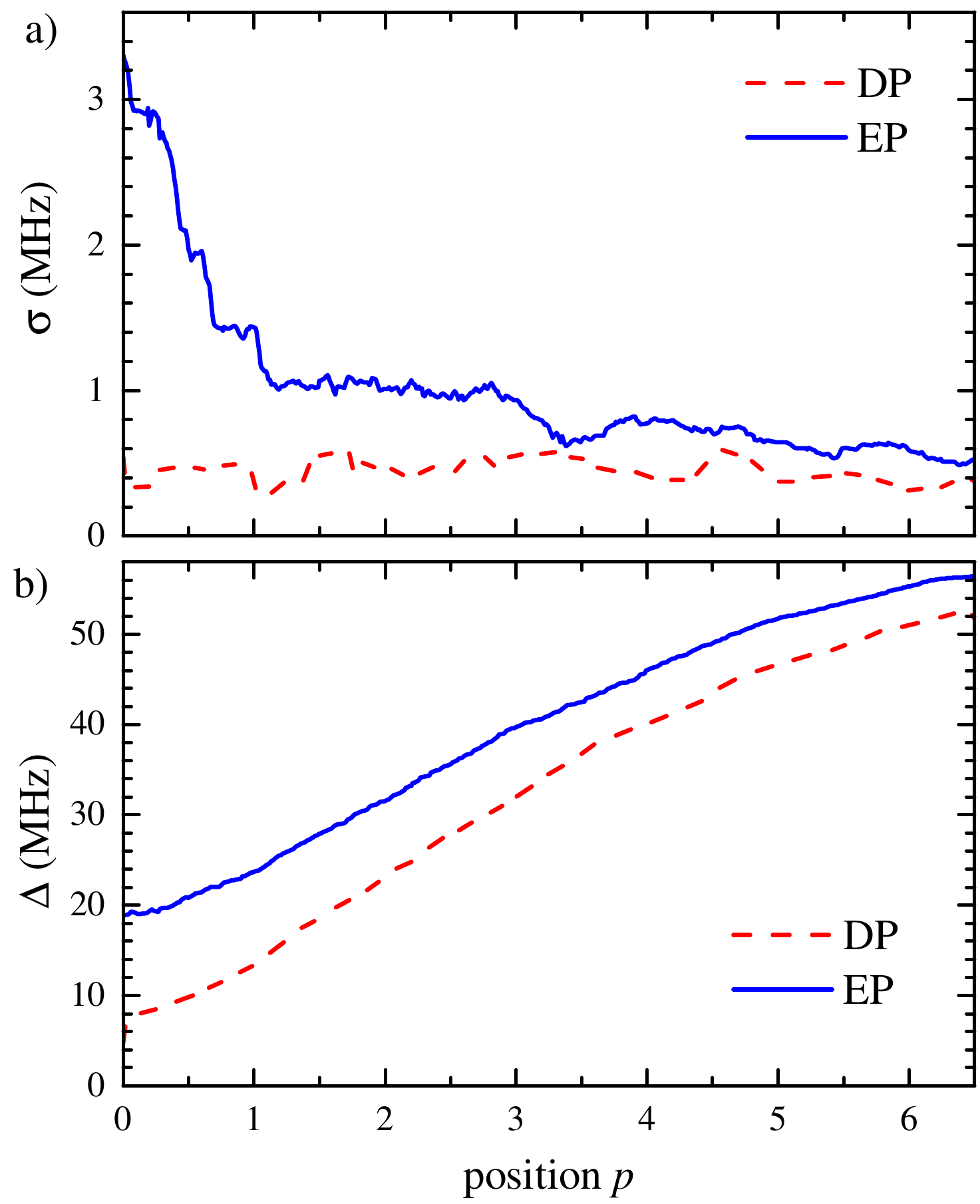}
	\caption{Standard deviation $\sigma$ (a) and magnitude (b) of the complex frequency splitting $\Delta$, for DP and EP sensors, as labelled.}
	\label{fig:Delta}
\end{figure}

We then average the square deviation using a 20-point adjacent averaging, to obtain the mean square deviation $\sigma^2_\gamma$ and $\sigma^2_\omega$, as function of $p$, for $\Delta_\gamma$ and $\Delta_\omega$, respectively. We then calculate the complex frequency splitting magnitude 
\be \Delta=\sqrt{\Delta_\omega^2+\Delta_\gamma^2} \ee 
and its mean square deviation
\be \sigma=\sqrt{\left(\frac{\sigma_\omega \Delta_\omega}{\Delta}\right)^2+\left(\frac{\sigma_\gamma \Delta_\gamma}{\Delta}\right)^2}\ee 
which are shown in \Fig{fig:Delta}. We can see that for the DP sensor, $\sigma$ is rather independent of $p$, indicating that for the DP sensor the frequency splitting is a good measure for the sensor precision. For the EP sensor instead, $\sigma$ significantly increases with decreasing splitting, indicating that for EP sensors the frequency splitting is not a good measure for the sensor precision.  

\begin{figure}
	\includegraphics[width=\columnwidth]{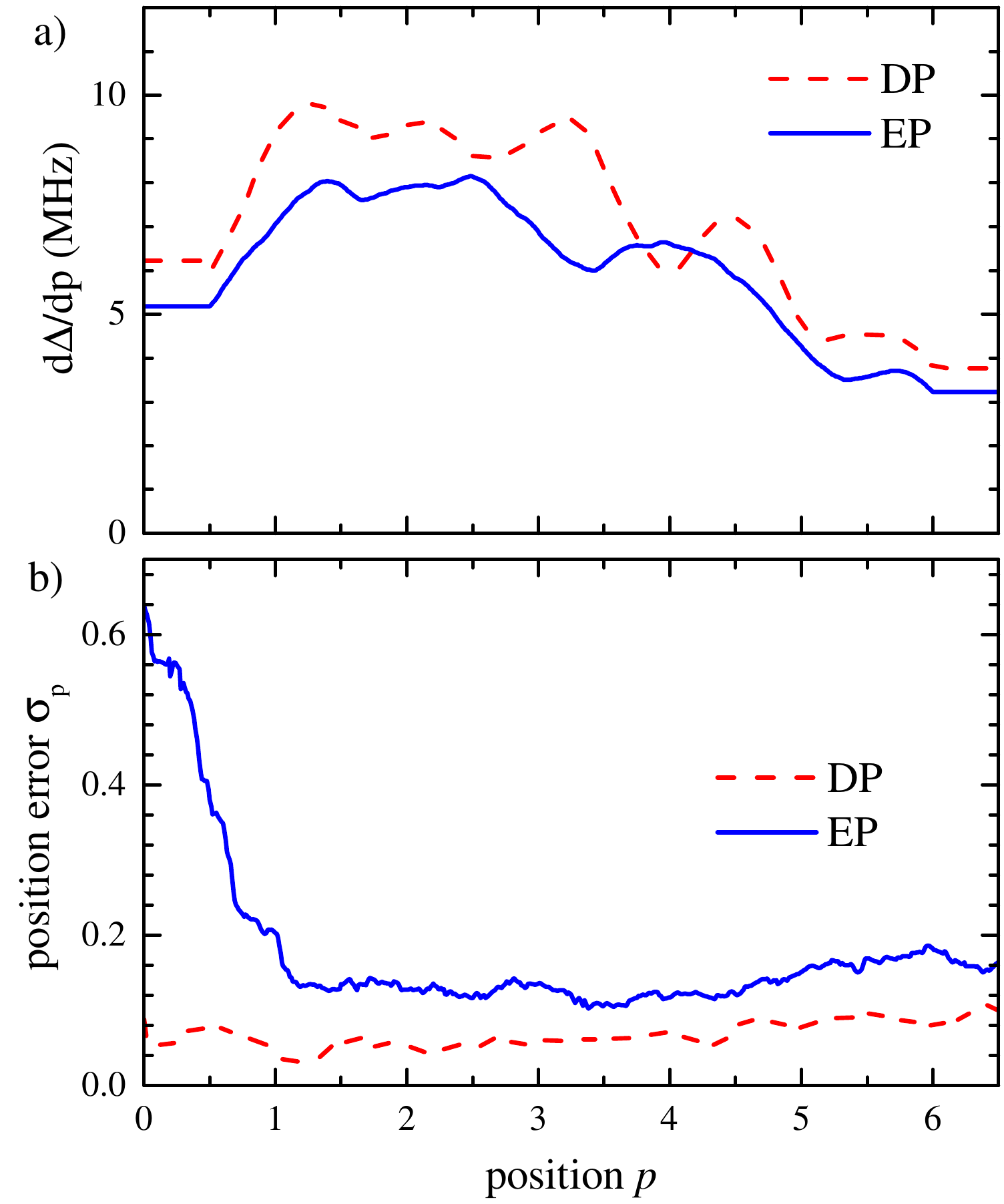}
	\caption{(a) $d\Delta/dp$, the change of the complex frequency splitting with $p$. (b) Calculated error $\sigma_p$ of the scatterer position measurement, for DP and EP sensors.}
	\label{fig:Precision}
\end{figure}

To calculate the rms error of the position $\sigma_p$, determined by $\Delta$, we use
\be \sigma_p = \frac{dp}{d\Delta}\sigma\ee
with the results shown in \Fig{fig:Precision}. The derivative was smoothed using the Savatski-Golay method with first order and 100 points to reduce its noise.

We see that the error in determining the position of the target scatterer for the EP sensor is a factor of 1.8 to 10 larger than for the DP sensor. We can therefore conclude that the data presented in Fig.2a of \cite{ChenN17} show a lower precision 
of the EP sensor as compared with the DP sensor. Importantly, the precision reduces for smaller perturbations, opposite to the behaviour expected from the scaling of the complex frequency splitting.

For the data analysis the program Origin 2016 (OriginLab, USA) was used.


\begin{thebibliography}{10}
	
	\bibitem{WiersigPRL14}
	J. Wiersig, Phys. Rev. Lett. {\bf 112},  203901  (2014).
	
	\bibitem{ZhangSR15}
	N. Zhang {\it et~al.}, Sci. Rep. {\bf 5},  11912  (2015).
	
	\bibitem{WiersigPRA16}
	J. Wiersig, Phys. Rev. A {\bf 93},  033809  (2016).
	
	\bibitem{RenOL17}
	J. Ren {\it et~al.}, Opt. Lett. {\bf 42},  1556  (2017).
	
	\bibitem{ChenN17}
	W. Chen {\it et~al.}, Nature {\bf 548},  192  (2017).
	
	\bibitem{HodaeiN17}
	H. Hodaei {\it et~al.}, Nature {\bf 548},  187  (2017).
	
	\bibitem{DoostPRA14}
	M.~B. Doost, W. Langbein, and E.~A. Muljarov, Phys. Rev. A {\bf 90},  013834
	(2014).
	
	\bibitem{HeissJPA12}
	W.~D. Heiss, J. Phys. A: Math. Theo. {\bf 45},  444016  (2012).
	
	\bibitem{DemtroderBook98}
	W. Demtr{\"o}der, {\em Laser Spectroscopy} (Spinger-Verlag, Berlin, 1998).
	
	\bibitem{BraginskyBook92}
	V.~B. Braginsky and F.~Y. Khalili, {\em Quantum Measurement} (Cambridge
	University Press, Cambridge, UK, 1992).
	
\end{thebibliography}

\end{document}